\documentclass[english,amsmath,amssymb,amsfonts,fleqn,10pt,floatfix,aps,superscriptaddress,twocolumn,prl]{revtex4-1}
\PassOptionsToPackage{usenames,dvipsnames}{xcolor}
\usepackage{pdfpages}
\usepackage{placeins}
\usepackage{enumitem}
\usepackage{color}
\usepackage{hyperref}
\hypersetup{colorlinks,
linkcolor=Red,
citecolor=OrangeRed,
urlcolor=BlueViolet,
anchorcolor=OliveGreen}
\usepackage{graphicx}
\usepackage{booktabs}
\usepackage{relsize}
\usepackage{stackrel}
\usepackage{units}
\usepackage{braket}
\usepackage{mathrsfs}
\usepackage{transparent}
\usepackage{verbatim}
\usepackage[us,12hr]{datetime}
\usepackage{bm}
\usepackage{bbm}
\usepackage{dsfont}

\usepackage{colordvi}

\newcommand{\mathsym}[1]{{}}
\newcommand{\unicode}[1]{{}}

\newcommand{\hide}[1]{}

\newcommand{\eq}[1]{Eq.\,(\ref{#1})}
\newcommand{\eqs}[1]{Eqs.\,(\ref{#1})}

\newcommand{\fig}[1]{Fig.\,\ref{#1}}

\newcommand{\brackets}[1]{\lbrace{#1\rbrace}}

\newcommand*{\dyad}[2]{ {\ket{#1}\hspace{-0.5ex}\bra{#2}}}
\newcommand*{\pdyad}[2]{ {\left(\ket{#1}\hspace{-0.5ex}\bra{#2}\right)}}

\newcommand{\bn}[0]{\mathbf{n}}
\newcommand{\by}[0]{\mathbf{y}}
\newcommand{\bx}[0]{\mathbf{x}}
\newcommand{\bchi}[0]{\bm{\chi}}
\newcommand{\ot}[0]{\mathrel{\hspace{-0.3ex}\scriptstyle{\otimes}\hspace{-0.3ex}}}
\newcommand{\mindyad}[2]{\mathrel{\scriptstyle{\dyad{#1}{#2}}}}
\newcommand{\da}[0]{\mindyad{0}{0}}
\newcommand{\db}[0]{\mindyad{0}{1}}
\newcommand{\dc}[0]{\mindyad{1}{0}}
\newcommand{\dd}[0]{\mindyad{1}{1}}
\newcommand{\op}[4]{#1 \ot #2 \ot #3 \ot #4}

\newcommand{\D}{\mathrm{d}}  

\newcommand{\brho}[0]{\bm{\mbox{\large \(\varrho\)}}}

\newcommand{\Or}{\mathcal{O}}

\newcommand {\ic}[0]{\ensuremath{\dot{\imath}}}
\newcommand{\ox}[2]{{[  \vcenter{\hbox{\(\displaystyle \mathop{{\scriptscriptstyle \chi}}\limits_{#1}^{#2} \)}}] }}

\begin{document}
\setcounter{section}{0}

\title{Certifying Separability in Symmetric Mixed States, and Superradiance}
\author{Elie Wolfe}
\email{wolfe@phys.uconn.edu}
\affiliation{Department of Physics, University of Connecticut, Storrs, CT 06269}

\author{S.F. Yelin}
\affiliation{Department of Physics, University of Connecticut, Storrs, CT 06269}
\affiliation{ITAMP, Harvard-Smithsonian Center for Astrophysics, Cambridge MA 02138}

\date{\today}

\begin{abstract}
Separability criteria are typically of the necessary-but-not-sufficient variety, in that satisfying some separability criterion, such as positivity of eigenvalues under partial transpose, does not strictly imply separability.  Certifying separability amounts proving the existence of a decomposition of target mixed state into some convex combination of separable states; determining the existence of such a decomposition is ``hard". We show that it is effective to instead ask if the target mixed state ``fits" some preconstructed separable form, in that one can generate a sufficient separability criterion relevant to all target states in some family by ensuring enough degrees of freedom in the preconstructed separable form. We demonstrate this technique by inducing a sufficient criterion for ``diagonally symmetric" states of N qubits. A sufficient separability criterion opens the door to study precisely how entanglement is (not) formed; we use ours to prove that, counter-intuitively, entanglement is not generated in idealized Dicke Model superradiance despite its exemplification of many-body effects. We introduce a quantification of the extent to which a given preconstructed parametrization comprises the set of all separable states; for ``diagonally symmetric" states our preconstruction is shown to be fully complete. This implies that our criterion is necessary in addition to sufficient, among other ramifications which we explore.
\end{abstract}
\maketitle

Despite extensive interest in many-body entanglement \cite{multireview,characterizingentanglement,entang.review.toth,NilpotentReview} the longstanding question of how, exactly, entanglement is generated {\em at all} remains open. To establish the minimal requisite common features of entanglement generation we must seek counter-intuitive instances to challenge our preconceptions. To that end, this research was motivated by initial indications which - inconclusively - suggested that entanglement may not be a feature of Dicke Model superradiance. Superradiance is a coherent radiative phenomenon resulting from collective and cooperative atomic effects \cite{superrad.original,superrad.yelin,*superrad.yelinBook,superrad2010}\footnote{Superradiance can also be understood as originating from superexchange, i.e., excitation-swapping between particles, or, equivalently, the virtual exchange of photons. Note, however, that the real exchange of photons leads to near- and far-field dipole-dipole interactions which lead to dephasing and the occupation of manifolds with lower symmetry.}, and thus it possesses the typical hallmark of an entangling process; see, for example \cite{NilpotentCavity}. Various necessary criteria for separability \cite{ppt.peres,ppt.horodecki,SymmetricEquivalents} nevertheless failed to find signatures of entanglement. The extraordinary claim ``superradiance occurs without entanglement", demands the highest standard of evidence; to prove that superradiance need not be entangling we must certify its separability by employing some sufficient separability criterion.

For pure states, various methods can be employed to quantify entanglement \cite{characterizingentanglement,entang.review.toth,NilpotentReview}. Mixed states, however, lack a general solution \cite{decomposition.cirac,threequbits}. Inspired in part by the generalization of Glauber-Sudarshan P invoked in Eq. (28) of Ref. \cite{decomposition.cirac}, we derived a separable decomposition applicable to superradiating systems. Whereas Ref. \cite{decomposition.cirac} is an existence proof, our decomposition explicitly solves a separability ansatz. Indeed, the bulk of our research effort was dedicated to identifying this sufficient separability criterion. Rewardingly, we subsequently realized that the technique we developed is applicable to far more than just superradiating systems; our approach for certifying separability is remarkably efficient throughout a broad class of states.

Our procedure amounts to explicitly parametrizing both the general family of states of interest, as well as some set of preconstructed separable states. Testing if the general-family parameters can be mapped to the separable-set parameters (``Does it fit?") is therefore a sufficient determination of separability. We demonstrate this method in detail on the ``General Diagonal Symmetric" states, within which Dicke Model superradiance evolves, and we successfully certify the perpetual separability of that model. This scenario is further exemplary in that our parametrization of separable states surprisingly appears to encompass {\em all} separable diagonally-symmetric states; thus the separability criterion developed in this paper is apparently not only sufficient but also necessary. 

We define the general diagonal symmetric (GDS) mixed states as those which are diagonal in the symmetric eigenbasis of $N$-partite $2$-level Dicke states. Each Dicke-basis pure state is a superposition of equal-energy states; it is the normalized sum-over-all-permutations of a (separable) computational-basis state. Using bold font to indicate sets, such as $\bn=\brackets{n_0,n_1}$, we have
\begin{align}\label{eq:firstdef}
   \ket{D_\bn} &=w_\bn \sum_{\begin{array}{c}\scriptstyle{\text{perms.}}\\ \scriptstyle{\brackets{\ket{0},\ket{1}}}\end{array}}{\ket{\underbrace{0...0}_{n_0},\underbrace{1...1}_{n_1}}}
\end{align}
where $n_0+n_1=N$ and $w_\bn=\sqrt{n_0!n_1!/N!}$\..

So for example
\begin{align}\label{eq:exampleket}
\ket{D_{3,1}}=\frac{\ket{0001}+\ket{0010}+\ket{0100}+\ket{1000}}{\sqrt{4}}.
\end{align}
The state $\ket{D_\bn}$ is entangled for all $0<n_0<N$; Dicke states are natural generalizations of the W state \cite{3Qubits2Ways}, and can also be described as the simultaneous eigenstates of total spin and spin-$z$ operators with $J=N/2$ and $M=\left(n_1-n_0\right)/2$.

The most general mixed state which is diagonal in this basis can be parametrized as
\begin{align}\label{eq:gds}
    \rho_{\text{\tiny GDS}}=\sum\limits_{\bn}^{}{{\chi_\bn} \ket{D_\bn}\bra{D_\bn}}
\end{align}
where the $\chi_\bn$ represent the eigenvalues in the eigendecomposition of $\rho_{\text{\tiny GDS}}$, which, in the convention of quantum optics, we refer to as the populations of $\rho_{\text{\tiny GDS}}$.

Next we preconstruct a set of separable states to serve as targets for our decomposition. We start with a completely generic single qubit pure state $\ket{\psi}=\sqrt{y} \ket{0}+\sqrt{1-y} e^{\ic\phi}\ket{1}$, defined as $\rho^1\left[y,\phi \right]\equiv\ket{\psi}\bra{\psi}$ in operator form, where we take an $N$-fold tensor product of the single qubit state with itself, and mix uniformly overl all phases but discretely over arbitrary amplitudes $y_j$ with weights $x_j$,
\begin{align}\label{eq.qubitSDSrawdef}
    \rho_{\text{\tiny SDS}}&\equiv  \int\limits_{0}^{2\pi}{ {\left(2 \pi\right)}^{-1}\sum\limits_{j=1}^{j_{\text{max}}}{x_j {{\left({\rho^1}\left[y_j,\phi\right]\right)}^{\otimes N}}\mbox{d}\phi}}\,.
\end{align}    
We call such parametrized states separable diagonally symmetric (SDS) states,, and the value of $j_{\text{max}}$ depends on $N$. Note that, by definition, all the variables $x_j$, $y_j$ appearing in \eq{eq.qubitSDSrawdef} must be real numbers between $0$ and $1$. Note also that our mixing protocol differs markedly from the Spherical Harmonics basis suggested in Ref. \cite{decomposition.cirac}, and furthermore, the SDS states cannot be resolved by the partial-separability method of Ref. \cite{KrausCiracKarnasLewenstein2000}, as that protocol is incompatible with continuous mixtures. 

As proven in the supplementary online materials, \eq{eq.qubitSDSrawdef} can be equivalently expressed as
\begin{align}\label{eq.qubitSDS}
    \rho_{\text{\tiny SDS}}&=N!\sum\limits_{\bn}^{}{\sum\limits_{j=1}^{j_{\text{max}}}{{\frac{x_j {y_j}^{n_0}{(1-y_j)}^{n_1}}{n_0!n_1!}}\ket{D_\bn}\bra{D_\bn}}}
\end{align}	
which more clearly parallels the form of \eq{eq:gds}. Orthogonality of the Dicke states allows us to match up terms inside the sums of \eq{eq:gds} and \eq{eq.qubitSDS}, implying $N+1$ polynomial equations \footnote{There are $N+1$ ways to choose $\bn$, since $n_0,n_1 \in \mathbb{Z}^+$ and $n_0+n_1=N$.} which define a decomposition the populations $\bchi$ of $\rho_{\text{\tiny GDS}}$ into the parameters $\bx,\by$ of a $\rho_{\text{\tiny SDS}}$. Explicitly, if we can successfully identify a mapping
\begin{align}\label{eq.qubitdecomp}
	\forall_\bn \quad {\chi_\bn} \,=\, N!\sum_{j=1}^{j_{\text{max}}}{\frac{x_j {y_j}^{n_0}\left(1-{y_j}\right)^{n_1}}{n_0!n_1!}}
\end{align}
then we will have demonstrated that our particular $\rho_{\text{\tiny GDS}}$ exists in the subspace defined by all possible $\rho_{\text{\tiny SDS}}$, $\rho_{\text{\tiny GDS}} \in \brho_{\text{\tiny SDS}}$, and thus that $\rho_{\text{\tiny GDS}}$ is necessarily separable.

$j_{\text{max}}$ is chosen in order for the system of equations (\ref{eq.qubitdecomp}) to be well behaved, i.e. that there should be exactly $N+1$ variables $\bx,\by$ appearing in the $N+1$ equations. Considering that $x_j$ and $y_j$ always come in pairs then plainly when $N+1$ is even we should set $j_{\text{max}}=\left(N+1\right)/2$. When $N+1$ is odd the situation requires a manual adjustment; we take $j_{\text{max}}=\lceil\left(N+1\right)/2\rceil$ and fix the extraneous variable by forcing $y_{\left(N+2\right)/2}=0$ \footnote{Setting $y_{\left(N+2\right)/2}=0$ when $N$ is an even number actually doesn't induce loss of generality, evidenced in that $\operatorname{PPTGDSVol}=\operatorname{SDSVol}$ still holds when $N$ is an even number.}. To demonstrate, here is the  system of polynomial equations for $N=4$ qubits,
\begin{align}\label{eq:explicit4}
    \hspace{-3ex}\begin{array}{lll}
    \chi_{4,0}&=& {x_{1}{{{({y_{1}})}}}^{4}+x_{2}{{{({y_{2}})}}}^{4}} 
    \\\chi_{3,1}&=& 4\left({x_{1}{{{({y_{1}})}}}^{3}{({1-y_{1}})}+x_{2}{{{({y_{2}})}}}^{3}{({1-y_{2}})}} \right)
    \\\chi_{2,2}&=& 6\left({x_{1}{{{({y_{1}})}}}^{2}{{{({1-y_{1}})}}}^{2}+x_{2}{{{({y_{2}})}}}^{2}{{{({1-y_{2}})}}}^{2}} \right)
    \\\chi_{1,3}&=& 4\left({x_{1}{({y_{1}})}{{{({1-y_{1}})}}}^{3}+x_{2}{({y_{2}})}{{{({1-y_{2}})}}}^{3}} \right)
    \\\chi_{0,4}&=& {x_{1}{{{({1-y_{1}})}}}^{4}+x_{2}{{{({1-y_{2}})}}}^{4}+x_{3}} \quad\text{\footnote{Note that $x_3$ does not appear until the final equation, this is a consequence of having set $y_3=0$ to ensure that only five free variables exist in the five equations (\ref{eq:explicit4}).}}
    \end{array}
\end{align}
Importantly, although the system of equations  mapping $\bchi \Leftrightarrow \bx,\by$ can always be solved, the decomposition is valid only if it passes a ``sanity check". Explicitly, {\em this} decomposition certifies that $\rho_{\text{\tiny GDS}}$ is separable if and only if convexity conditions on the coefficients parametrizing $\rho_{\text{\tiny SDS}}$ are satisfied,
\begin{align}\label{eq.sepcrit}\begin{split}
    &\rho_{\text{\tiny GDS}}\in\brho_{\text{\tiny SDS}}\;\text{ iff }\;\exists\;{\bx,\by}\text{ satisfying \eq{eq.qubitdecomp}}
    \\&\text{such that }\;\forall_{j}:\; 0\le \,x_j\,,\,y_j\,\le 1\quad\text{\footnote{$\sum{x_j}=\mbox{Tr}\left[\rho_{\text{\tiny GDS}}\right]=1$ is guaranteed by normalization.}.}
\end{split}\end{align}
To be clear, conditions (\ref{eq.sepcrit}) are cumulatively a sufficient criterion for certifying separability, since 
\begin{align}
&\brho_{\mbox{\tiny SDS}}~\subseteq~\brho_{\mbox{\tiny SEP}\cap\mbox{\tiny GDS}}~\subset~\brho_{\mbox{\tiny GDS}}
\\\nonumber &\text{where }\;\brho_{\mbox{\tiny SEP}\cap\mbox{\tiny GDS}} \equiv \brho_{\mbox{\tiny SEP }}\cap\brho_{\mbox{\tiny GDS}}
\end{align}
and where $\subseteq$ and $\subset$ are analogous to $\leq$ and $<$ respectively; $\subset$ indicates a {\em proper} subset, categorically rejecting the possibility of equivalence. So, even though we have not yet ruled out the existence of a separable $\rho_{\text{\tiny GDS}}$ incompatible with the SDS format, the criterion developed is already a sufficient one.

The ability to certify full separability is highly desired, as: 
\vspace{-1ex}\begin{enumerate}
\item The necessary separability criterion of positivity under all partial transpositions \cite{ppt.peres,ppt.horodecki} does {\em not} imply biseparability along all bipartitions \cite{upb.original,acin2001bound}.
\vspace{-3ex}\item A state can be partially separable, e.g. separable along all bipartitions, but {\em still be entangled} \cite{Szalay2012Partial}, even to the extent of serving as a resource for Bell inequality violations \cite{ppt.nonlocality}.
\end{enumerate}
\vspace{-1ex}We emphasize that this method of generating sufficient (full) separability criteria is generic and adaptable: developing criteria for different states means parametrizing some separable states of similar form, so as to allow for parameter matching. 

To demonstrate the utility of possessing a sufficient separability criterion we assess the candidacy of superradiance for entanglement generation, per the original motivation for this research. A system initially in a pure Dicke state is said to evolve according to idealized pure Dicke Model superradiance \cite{superrad.original} if it decays to the ground state according to the first-order differential equations
\begin{align}\label{eq:superrad}\begin{split}
    \hspace{-2ex}\frac{\partial \chi_{n_0,n_1}\left[\tau\right]}{\partial \tau}\;=\;\begin{array}{l}
    -{\left(n_0+1\right)n_1 \chi_{n_0,n_1}\left[\tau\right]}
    \\+{n_0\left(n_1+1\right)\chi_{n_0-1,n_1+1}\left[\tau\right]}
    \end{array}
\end{split}\end{align}
where $\tau$ is a dimensionless time parameter, $\tau=\Gamma~t$ \footnote{\eq{eq:superrad} corresponds to Eq. (4.7) in Ref. \cite{superrad.original}, but with $\:t~\times~\Gamma~\rightarrow~\tau$, $J~-~M~\rightarrow ~n_0$, $J~+~M~\rightarrow~n_1$, and using $\rho_M~\rightarrow~\chi_{n_0,n_1}$ for the populations}. The idealization is that of perfect indistinguishability of the particles; experimentally it corresponds to the small-volume limit without dipole-dipole induced dephasing. Our question is whether such idealized superradiance can generate entanglement.

Intuitively, this indistinguishable-particles idealization should yield the {\em strongest} entanglement possible, such that if {\em less}-idealized superradiance were to generate entanglement, then presumably entanglement would also be evident in this extremal model; see for example the discussion of volume-dependent many-body effects in Ref. \cite{superrad.yelin,*superrad.yelinBook}\footnote{Note that this presumption does not constitute proof; we cannot confidently infer an absence of entanglement in the realistic cases of dephasing and lower symmetry from our null finding of entanglement in the pure Dick Model. Proof of inference is desirable for future research.}. To consider entanglement generation we utilize an unentangled initial state; the only non-ground, separable, pure, Dicke state, is the maximally excited state \footnote{Alternative separable initial states include the SDS states (which are not pure), pure superpositions of Dicke states, and even mixed states outside of the GDS manifold. The authors consider such variants of initial conditions, along with other generalizations of superradiance, in another paper now in preparation.}, i.e. we use initial conditions 
\begin{align}\label{eq:initconds}
	\chi_\bn \left[\tau\rightarrow 0\right]=\begin{cases}1& {n_1=N,n_0=0}\\ 0& {n_1<N,n_0>0}\end{cases} \,. 
\end{align}
Solving the differential equations yields populations $\bchi$ as functions of $\tau$; one may then test the system for separability at any time $\tau$. Consider the Peres-Horodecki criterion \cite{ppt.peres,ppt.horodecki}, which notes that genuinely separable states remain positive-semidefinite under partial transpositions (PPT). The property of PPT is necessary but insufficient for separability \cite{upb.original,acin2001bound,Szalay2012Partial,ppt.nonlocality}, although for symmetric states it {\em is} sufficient for $N=2,3$, but still insufficient for $N\geq 4$ \cite{eckert2002quantum,ppt4qubit,Lowenstein2012PPT}. We find that the PPT criterion is satisfied for all $\tau>0$ for all $N\leq 10$ \footnote{PPT of superradiance may be demonstrated, for example, by graphing the eigenvalues of the partial transpositions of superradiant $\rho$ as functions of time, analogous to the graphs of the decomposition parameters in the supplementary online materials.}. This consistency-with-separability per the PPT criterion underscores the need for an unambiguous, i.e. sufficient, criterion, a challenge which conditions (\ref{eq.sepcrit}) rise to fulfill.

To certify separability one merely inspects the decomposition parameters $\brackets{\vec{x},\vec{y}}$ obtained by substituting the solved-for populations $\bchi_\bn[\tau]$ into the system of polynomial equations given by \eq{eq.qubitdecomp}. Certification amounts to verification that $\brackets{\vec{x},\vec{y}}$ satisfy conditions (\ref{eq.sepcrit}). Indeed, we numerically verified that for pure Dicke Model superradiance, conditions (\ref{eq.sepcrit}) are satisfied for all $\tau>0$, thereby certifying full separability throughout the time evolution, for $N\leq 8$. This is demonstrated graphically in the supplementary online materials for both $N=4$ and $N=8$.


We now conjecture that {\em whenever} a state $\rho_{\text{\tiny GDS}}$ is entangled then conditions (\ref{eq.sepcrit}) {\em must} be violated, making conditions (\ref{eq.sepcrit}) a {\em necessary and sufficient} separability criterion. The sufficiency is by construction, the necessity we can demonstrate by comparison to a known necessary criterion, namely PPT \cite{ppt.peres,ppt.horodecki}. We evidence that, upon restricting to GDS states, the PPT criterion coincides with conditions (\ref{eq.sepcrit}). We claim
\begin{align}
\label{PPTSEP}&\text{Lemma: }\quad \brho_{\mbox{\tiny SDS}}=\brho_{\mbox{\tiny SEP}\cap\mbox{\tiny GDS}}=\brho_{\mbox{\tiny PPT}\cap\mbox{\tiny GDS}}
\end{align}
where we {\em prove} Lemma (\ref{PPTSEP}) for $N=4$ and conjecture that it continues to hold for all $N$ \footnote{States diagonal in the symmetric basis are a subset of general permutation-symmetric states, $\brho_{\mbox{\tiny GDS}}\subset \brho_{\mbox{\tiny SYM}}$, thus Lemma (\ref{PPTSEP}) is both trivially true for $N=2,3$ \cite{ppt4qubit,eckert2002quantum} and consistent with the existence of permutation-symmetric PPT-entangled states \cite{ppt4qubit,Lowenstein2012PPT}.}. Demonstrating Lemma (\ref{PPTSEP}) may seem rather daunting; proving equivalence between separability criteria with formal logic is indeed an intimidating task. However, we can skip the logical proof and instead use integration to directly establish that volume of both $\brho_{\mbox{\tiny SDS}}$ and $\brho_{\mbox{\tiny PPT}\cap \mbox{\tiny GDS}}$ are identical. To do so we establish a metric on the spaces of density matrices, the metric can be arbitrary but must be consistent; we choose the populations of $\rho_{\text{\tiny GDS}}$ as our integration coordinates \cite{statevolume}\footnote{To be clear, this metric is entirely unconventional. See Ref. \cite{statevolume} for canonical mixed state space volumes.}. Thus 
\begin{align}\label{GDSVolIntegral}
&\hspace{-\mathindent}\operatorname{PPTGDSVol}_{N=4}=\int\limits_0^1 \!\!\!\int\limits_0^1 \!\!\!\int\limits_0^1 \!\!\!\int\limits_0^1 \!\!\!\int\limits_0^1 \bm{1}_{\mbox{\tiny PPT}}{\left(\bchi\right)} \delta_{\left(1-{{|\bchi|}_{1}}\right)}\:\D\bchi 
\end{align}
where $\,{|\bchi|}_1={\displaystyle \sum\limits_{\bn}^{}{\chi_\bn}}\,$ and $\bm{1}_{\mbox{\tiny PPT}}{\left(\bchi\right)}=\begin{cases}1 & \!{\bchi \in \brho_{\mbox{\tiny PPT}}}\\ 0& \!{\bchi \not\in \brho_{\mbox{\tiny PPT}}}\end{cases}$ is an indicator function which cuts off the integration whenever the populations violate the PPT conditions. Here the PPT conditions mean that all eigenvalues are nonnegative {\em for all} bipartitions of the qubits for partial transposition \footnote{The permutation symmetry of $\rho_{\text{\tiny GDS}}$ means we need only consider two bipartitions: partial transposition of the first qubit $\rho^{\operatorname{PT}_{1|3}}$ or of the first two qubits $\rho^{\operatorname{PT}_{2|2}}$, akin to the considerations in Ref. \cite{ppt4qubit}. Both bipartitions are rejected by the PPT indicator function (and both are necessary). The PPT indicator function may be thought of as a multidimensional Heaviside step function of the eigenvalues of $\rho^{\operatorname{PT}_{1|3}}$ and $\rho^{\operatorname{PT}_{2|2}}$. The authors used a different but equivalent formulation when evaluating \eq{GDSVolIntegral}.}. We find numerically that $\operatorname{PPTGDSVol}_{N=4}=\left(3808\pm 2\right)\times {10}^{-6}$. In contrast, the volume of {\em all} GDS states, including entangled, follows from \eq{GDSVolIntegral} absent the indicator function; $\operatorname{GDSVol}_{N}=1/N!$. For four qubits $\operatorname{GDSVol}_{N=4}=41,666.\bar{6}\times {10}^{-6}$.

In principle one could calculate the volume of $\brho_{\mbox{\tiny SDS}}$ along the same lines as \eq{GDSVolIntegral} but with a different indicator function based on conditions (\ref{eq.sepcrit}), but there is a much easier way to do it: perform the integration for SDSVol using $\bx$ and $\by$ as the integration coordinates, thus eliminating the need for any indicator function whatsoever. To stay consistent to the originally established metric of the populations $\bchi$, we must insert a volume element in the integrand, namely the absolute value of the determinant of Jacobian matrix for the change-of-variable. For $N=4$ there are five $\chi_\bn$ expressible in terms of $\bx,\by$ via \eq{eq.qubitdecomp}, which correspond to the columns of the Jacobian matrix. The five rows of the Jacobian matrix are the given by taking the derivative of the $\bchi$ list with respect to each of $x_1,x_2,x_3,y_1,y_2$. The Jacobian's determinant, happily {\em a priori} nonnegative, is $jac=96\, x_1\, x_2 {(1-y_1)}^{2} {(1-y_2)}^{2} {(y_1-y_2)}^{4}$. Lastly we must ensure a one-to-one mapping between $\bchi$ and $\bx,\by$. To avoid the problematic interchangeability between the variable pairs $x_1,y_1$ and $x_2,y_2$ we impose the ordering $x_1 \geq x_2$. 

Therefore 
\begin{align*}
\hspace{-4ex}\operatorname{SDSVol}_{N=4}=\int\limits_0^1 \!\!\!\int\limits_0^1 \!\!\!\int\limits_0^1 \!\!\!\int\limits_0^1 \!\!\!\int\limits_0^1 \bm{1}_{x_1 {\scriptscriptstyle \geq} x_2}\times jac\times \delta_{\left(1-{{|\bx|}_{1}}\right)}\:\D \bx \D\by
\end{align*}
where ${|\bx|}_{1}=\sum_{k=1}^{3}{x_k}$, and unlike the $\bx$, the $\by$ variables have no further restrictions placed upon them due to the normalization of $\rho_{\mbox{\tiny SDS}}$. We find that $\operatorname{SDSVol}_{N=4}=2/525 \approx \left(3809.5\right)\times {10}^{-6}$. Because we {\em must} have $\brho_{\mbox{\tiny SDS}} \subseteq \brho_{\mbox{\tiny PPT}\cap\mbox{\tiny GDS}}$ we are forced to revise $\operatorname{PPTGDSVol}_{N=4}$ to the upper limit of its uncertainty, which indicates convincingly that Lemma (\ref{PPTSEP}) is true for $N=4$. 

The authors suspect that Lemma (\ref{PPTSEP}) is true for all $N$ for reasons as follows: As previously mentioned, we found that Dicke Model superradiance time evolution, per \eq{eq:superrad}, is PPT for any $\tau\geq 0$ for at least $N\leq 10$. Thus superradiance serves as a sort of representative sample of PPT$\cap$GDS states, or formally $\brho_{\mbox{\tiny SUP-RAD}}\subset \brho_{\mbox{\tiny PPT}\cap\mbox{\tiny GDS}}$. But also as mentioned earlier, we found that such systems apparently {\em always} fit the SDS form, in that they satisfy conditions (\ref{eq.sepcrit}) for any $\tau\geq 0$ for at least $N \leq 8$. If Lemma (\ref{PPTSEP}) were false, then the unflappable fitting of superradiant states into the SDS form would be surprising, as we would have expected $\brho_{\mbox{\tiny SUP-RAD}}\not\subset \brho_{\mbox{\tiny SDS}}$.  Thus we have accumulated evidence-by-contraposition to support Lemma (\ref{PPTSEP}) for $N>4$.

If Lemma (\ref{PPTSEP}) is true for all $N$, as evidence suggests, then the ramifications are numerous. First, it implies that conditions (\ref{eq.sepcrit}) amount to a {\em necessary} and sufficient criterion for separability. Second, it implies that the basic PPT criterion is a {\em sufficient} separability test for diagonally symmetric states. Third, we can generate novel practical necessary (but not sufficient) separability criteria by simply considering weaker extensions of conditions (\ref{eq.sepcrit}). For example, presuming that {\em all} separable diagonally symmetric states fit the form of \eq{eq.qubitdecomp} allows us to identify ''separable maxima" for the populations such that if even a {\em single} population exceeds its "maximum separable value" then entanglement is incontrovertible. We find that for $\rho_{\text{\tiny GDS}}$ to be separable it is necessary (but not sufficient) to satisfy this weaker form of \eq{eq.qubitdecomp} expressed as
\begin{align}
&\forall_\bn \quad \chi_{n_0,n_1}   \leq {\left(\frac{n_0!n_1!}{N!}\right)}^{-1} \operatorname*{max}_{0<y<1}{\left[{y}^{n_0}\left(1-{y}\right)^{n_1}\right]}\nonumber\\
	&\therefore \quad \chi_{n_0,n_1} \leq \left(\frac{{n_0}^{n_0} }{n_0!}\right)  \left(\frac{{n_1}^{n_1}}{n_1!}\right) \left(\frac{ N!}{N^N}\right)\,
\end{align}
which is computationally optimal as a first-pass test to detect entanglement. 

The symmetric basis of Dicke states can be extended to general qudits. We desire a generalization of \eq{eq.qubitdecomp} for qudits, and we wonder if said generalization would also be necessary in addition to sufficient, \'a la Lemma (\ref{PPTSEP}). We hope to consider this in a future work.

In conclusion, what was originally an analysis of superradiance has led to broad approach for studying multipartite entanglement. We found that a {\em Guess \& Check} technique can be surprisingly efficient, as evidenced by the derivation of conditions (\ref{eq.sepcrit}) which apply for all states diagonal in the symmetric basis. Moreover, the derived criterion is a completely tight characterization of separability properties, since we found that it maps out a volume of states no smaller than that defined by the PPT criterion. Additionally, our motivating question has been firmly answered in the negative; pure Dicke Model superradiance {\em cannot} generate entanglement, begging the question ''What {\em is}, then, the essential prerequisite of entanglement"? We hope that our techniques for generating sufficient separability criteria, and for certifying the sufficiency of known necessary separability criteria, may prove useful in furthering the understanding of entanglement. 

We thank Szil\'{a}rd Szalay of Budapest University for feedback on phrasing and notation. We wish to thank the NSF and the AFOSR for funding.

\bibliographystyle{apsrev4-1}
\bibliography{SuperRadCondensed}
\FloatBarrier
\onecolumngrid
\vspace{17ex}
\noindent\makebox[\linewidth]{\rule{\paperwidth}{1pt}} 
\appendix
\vspace{2ex}
\chapter{SUPPLEMENTARY ONLINE MATERIALS}

\renewcommand{\theequation}{A.\arabic{equation}}
\setcounter{equation}{0}

\noindent\makebox[\linewidth]{\rule{\paperwidth}{1pt}} 
\section{Explicit Separability Certification for N=4}

\begin{figure}[h]
\centering
\begin{minipage}[t]{.48\textwidth}
    \includegraphics[width=1.0\linewidth]{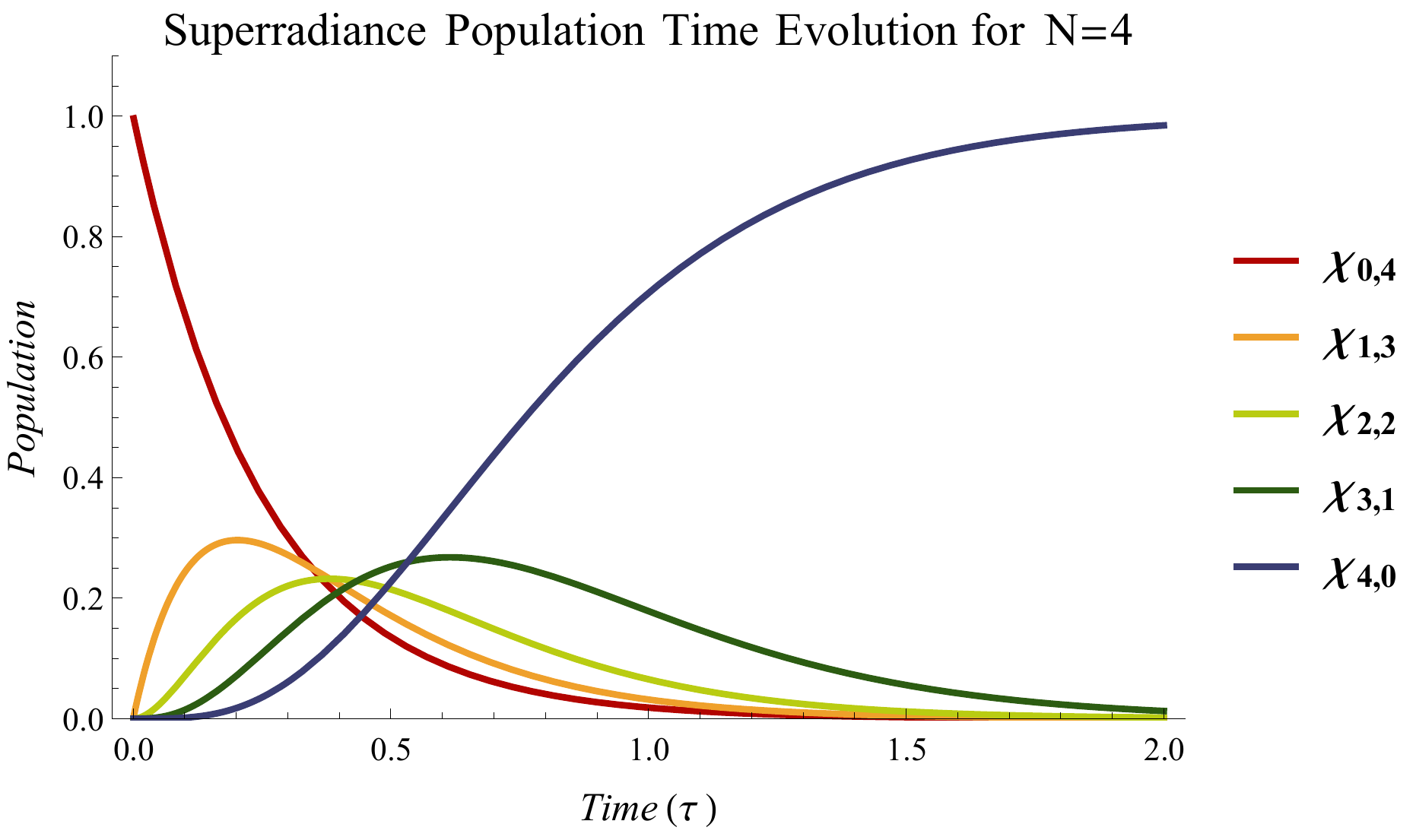}
    \caption{The system is initially entirely in the maximally-excited state, so the population $\chi _{0,4}$ initially equals 1. The system then cascades through the lower levels, such that the lower populations achieve their peak filling in chronological sequence, with the system asymptotically tending towards the ground state, defined by $\chi _{0,4}=1$. Observe that the sum of the five populations is equal to 1 at all times by virtue of normalization.}\label{fig:n4superad}
\end{minipage}\hfill
\begin{minipage}[t]{.48\textwidth}
    \includegraphics[width=1.0\linewidth]{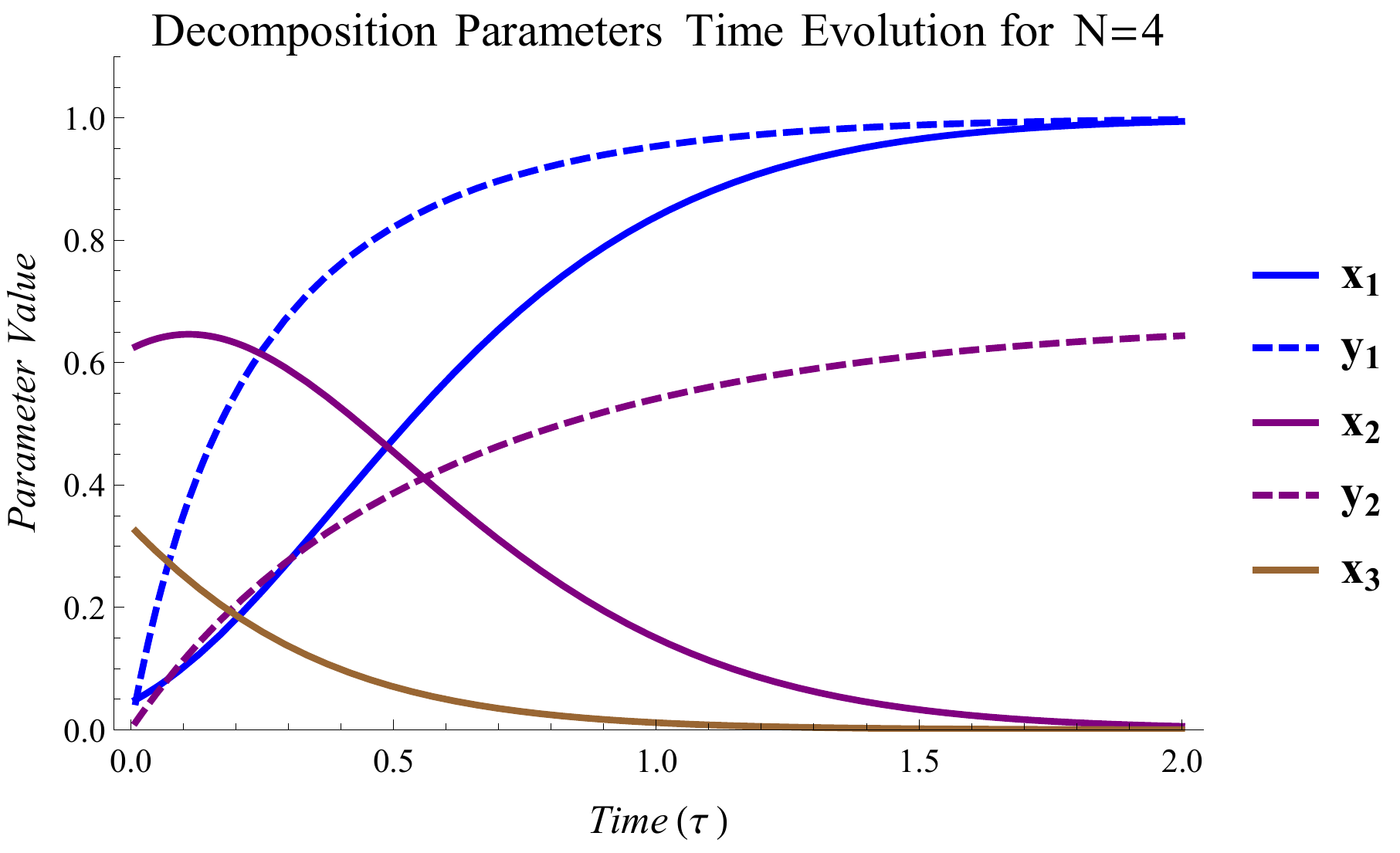}
    \caption{Observe that all five decomposition parameters remain bounded between zero and one, which is to say that conditions (\ref{eq.sepcrit}) of the main text are satisfied, and the system is perpetually fully separable. That the state is initially fully excited can be seen in that $x_2+x_3=1$ at $\tau=0$ and, although $y_3\equiv 0$, we see that $y_2$ {\em also} equals zero when $\tau=0$.  Normalization of the state imposes $x_1+x_2+x_3$=1 at all times.}\label{fig:n4decomp}
\end{minipage}
\end{figure}

In the main text we consider Dicke Model superradiance to be governed by the differential equations of \eq{eq:superrad} subject to initial conditions given by \eq{eq:initconds}, namely
\begin{align}\begin{split}\label{eq:superradSOM}
    \forall_\bn \quad & \frac{\partial \chi_{n_0,n_1}\left[\tau\right]}{\partial \tau}\;=\;
    {n_0\left(n_1+1\right)\chi_{n_0-1,n_1+1}\left[\tau\right]}-{\left(n_0+1\right)n_1 \chi_{n_0,n_1}\left[\tau\right]}     \\
    \text{such that}\quad&\chi_{n_0,n_1} \left[\tau\rightarrow 0\right]\;=\;\begin{cases}1& {n_1=N,n_0=0}\\ 0& {n_1<N,n_0>0}\end{cases} \,. 
\end{split}\end{align}
which for $N=4$ yields the solutions
\begin{align}\begin{split}\label{eq:4qubitsuperradSOM}
\chi _{0,4} = & e^{-4 \tau } \\
\chi _{1,3} = & 2 e^{-4 \tau }-2 e^{-6 \tau } \\
\chi _{2,2} = & 6 e^{-6 \tau } (-2 \tau -1)+6 e^{-4 \tau } \\
\chi _{3,1} = & 36 e^{-4 \tau } (\tau -1)+36 e^{-6 \tau } (\tau +1) \\
\chi _{4,0} = & e^{-6 \tau } (-24 \tau -28)+e^{-4 \tau } (27-36 \tau )+1
\end{split}\end{align}
which are plotted in \fig{fig:n4superad}.

Per \eq{eq.qubitdecomp} in the main text, the decomposition parameters are solved from the simultaneous polynomial equations defined by
\begin{align}\label{eq.qubitdecompSOM}
	\forall_\bn \quad {\chi_{n_0,n_1}} \,=\, N!\sum_{j=1}^{j_{\text{max}}}{\frac{x_j {y_j}^{n_0}\left(1-{y_j}\right)^{n_1}}{n_0!n_1!}}\,.
\end{align}
Enumerated explicitly for $N=4$ the decompositions equations are
\begin{align}\label{eq:explicit4SOM}
    \begin{array}{lll}
    \chi_{4,0}& =& {x_{1}{{{({y_{1}})}}}^{4}+x_{2}{{{({y_{2}})}}}^{4}} 
    \\ \chi_{3,1}& =& 4\left({x_{1}{{{({y_{1}})}}}^{3}{({1-y_{1}})}+x_{2}{{{({y_{2}})}}}^{3}{({1-y_{2}})}} \right)
    \\  \chi_{2,2}& =& 6\left({x_{1}{{{({y_{1}})}}}^{2}{{{({1-y_{1}})}}}^{2}+x_{2}{{{({y_{2}})}}}^{2}{{{({1-y_{2}})}}}^{2}} \right)
    \\ \chi_{1,3}& =& 4\left({x_{1}{({y_{1}})}{{{({1-y_{1}})}}}^{3}+x_{2}{({y_{2}})}{{{({1-y_{2}})}}}^{3}} \right)
    \\ \chi_{0,4}& =& {x_{1}{{{({1-y_{1}})}}}^{4}+x_{2}{{{({1-y_{2}})}}}^{4}+x_{3}}
    \end{array}
\end{align}
which also appear as \eq{eq:explicit4} in the main text. One can readily solve \eqs{eq:explicit4SOM} analytically. To express the solutions it is convenient to relabel $y_1 = y_+,\, x_1=x+,\, y_2=y_-,\, \text{and}\, x_2=x_-$ so that we may compactly state
\begin{align}\begin{split}\label{eq:compactchi}
&\hspace{-3ex}y_{\pm} = \frac{9 \ox{3}{1}^2-18 \ox{1}{3} \ox{4}{0}+3 \ox{2}{2} \left(\ox{3}{1}-8 \ox{4}{0}\right) \pm \sqrt{324 \ox{1}{3}^2 \ox{4}{0}^2+12 \ox{2}{2} \left(8 \ox{2}{2}^2-27 \ox{1}{3} \ox{3}{1}\right) \ox{4}{0}-27 \left(\ox{2}{2}^2-3 \ox{1}{3} \ox{3}{1}\right) \ox{3}{1}^2}}{4 \ox{2}{2}^2+6 \left(\ox{3}{1}-4 \ox{4}{0}\right) \ox{2}{2}+9 \ox{3}{1}^2-9 \ox{1}{3} \left(\ox{3}{1}+4 \ox{4}{0}\right)} \\
&\hspace{-3ex}x_{\pm} = \frac{{y_{\mp}}^2 \chi_{2,2}-6 {\left(y_{\mp}-1\right)}^2 \chi_{4,0}}{6 {y_{\pm}}^2 \left(y_{\pm}-y_{\mp}\right) \left(y_{\pm} \left(2 y_{\mp}-1\right)-y_{\mp}\right)} \\
&\hspace{-3ex}x_3 = 1-x_+-x_-
\end{split}\end{align}
where in \eq{eq:compactchi} we used $\ox{n_0}{n_1}$ as merely a horizontally compact form of $\chi_{n_0,n_1}$. Taking the populations to be as per \eqs{eq:4qubitsuperradSOM} and then plotting the decomposition parameters as functions $\tau$ we obtain \fig{fig:n4decomp} where it is plainly evident that the extrama of $\brackets{\vec{x}(\tau),\vec{y}(\tau)}$ lie between zero and one. We know that the superradiating systems starts off in a separable state (the maximally excited state) and that it tends to a separable state (the ground state) and so if there {\em were} entanglement generated then it would have to build and then dissipate. As such, we confidently establish {\em permanent} separability when we are able to bound the extrama of $\brackets{\vec{x}(\tau),\vec{y}(\tau)}$ as between zero and one. This visually certifies the perpetual separability of the system, and hence the inability of pure Dicke Model superradiance to generate entanglement. 
\FloatBarrier
\noindent\makebox[\linewidth]{\rule{\textwidth}{6pt}} 
\section{Explicit Separability Certification for N=8}
\begin{figure}[ht]
\centering
\begin{minipage}[t]{.48\textwidth}
    \includegraphics[width=1.0\linewidth]{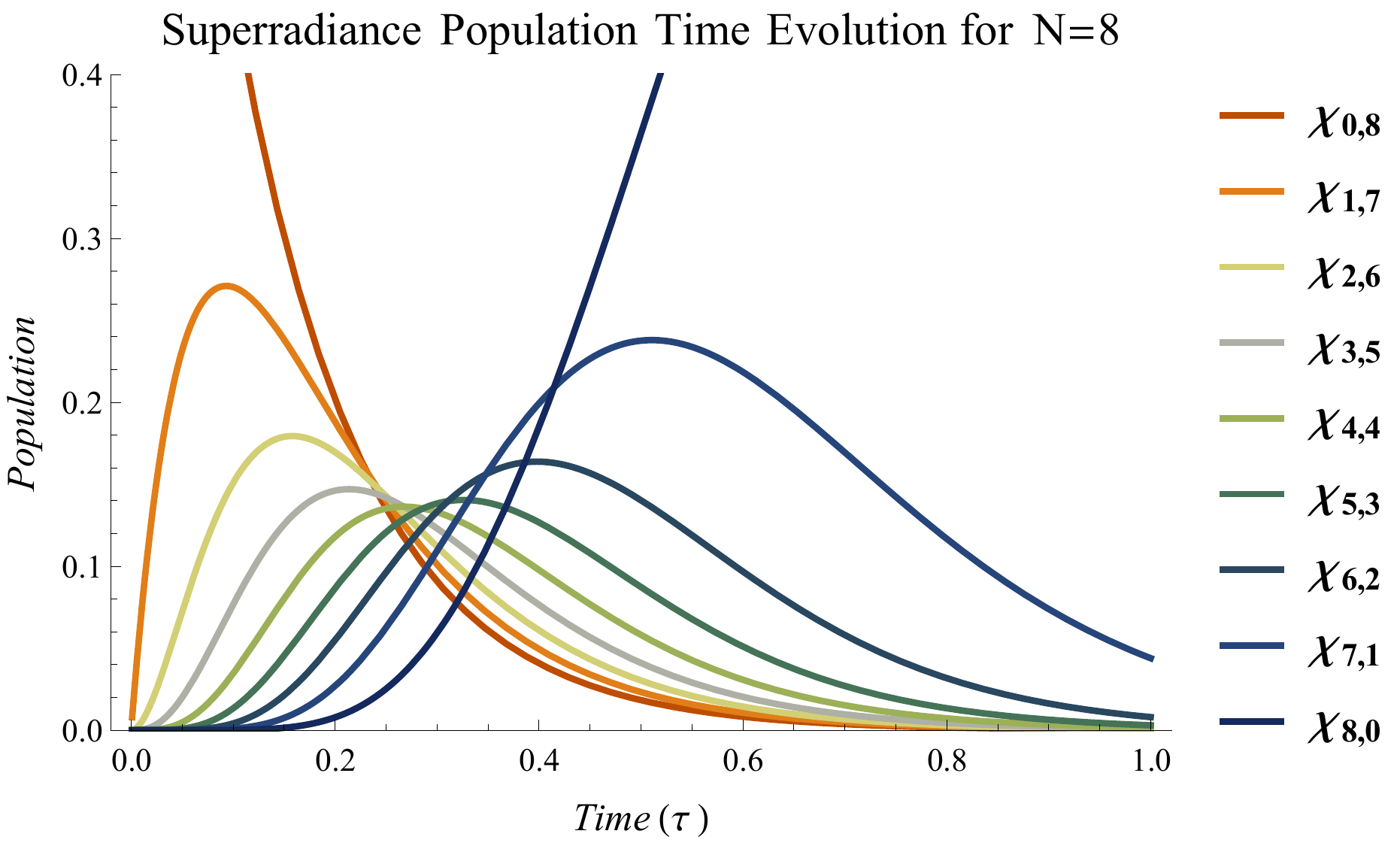}
    \caption{The state is initially entirely in the maximally-excited state, so the population $\chi _{0,8}$ initially equals 1. The system then cascades through the lower levels, such that the lower populations achieve their peak filling in chronological sequence, with the system asymptotically tending towards the ground state, defined by $\chi _{8,0}=1$. }\label{fig:n8superad}
\end{minipage}\hfill
\begin{minipage}[t]{.48\textwidth}
    \includegraphics[width=1.0\linewidth]{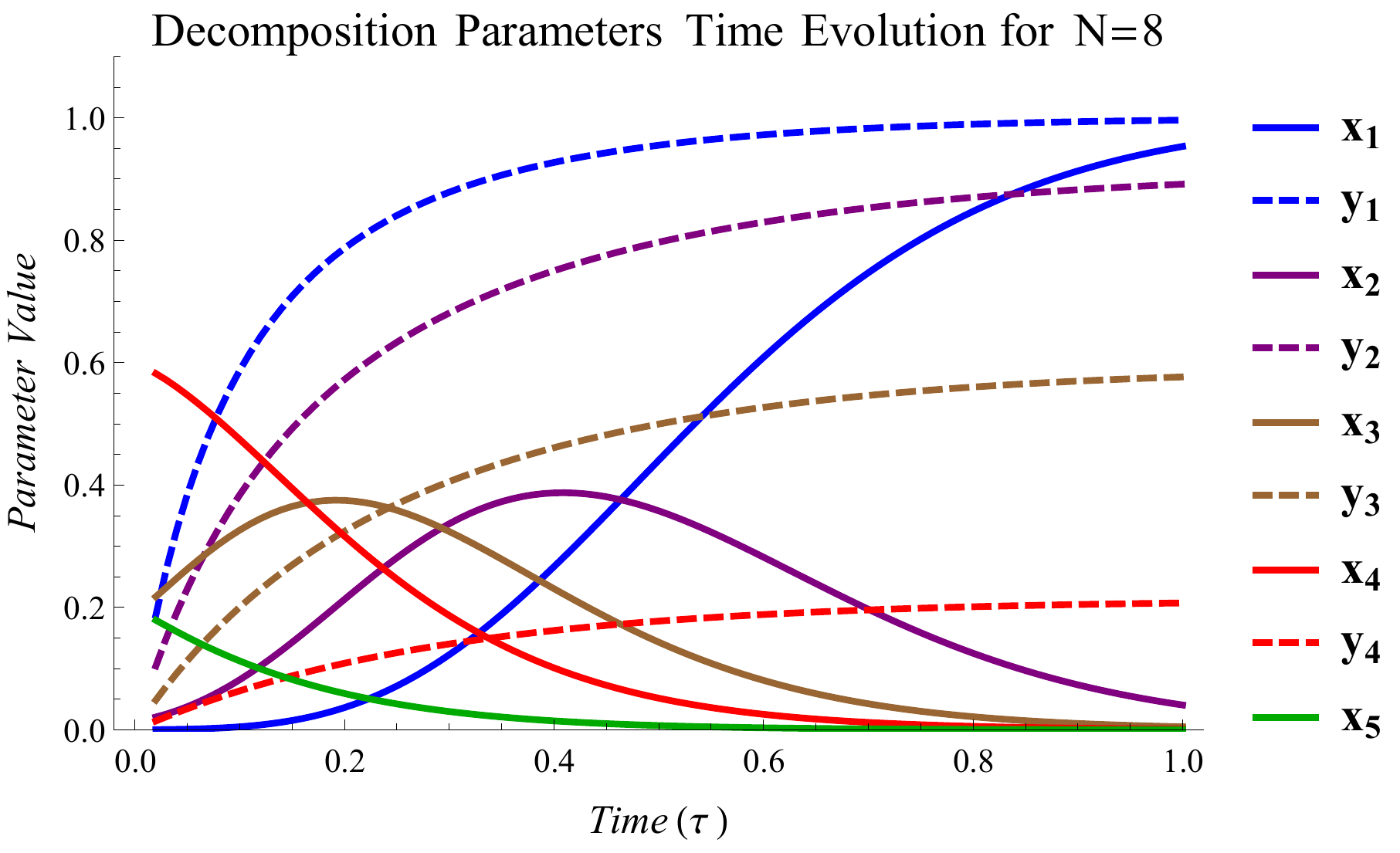}
    \caption{Observe that all nine decomposition parameters remain bounded between zero and one, which is to say that conditions (\ref{eq.sepcrit}) of the main text are satisfied, and the system is perpetually fully separable. Normalization is evident in that the sum of the $\bx$ totals one at all times $\tau$.}\label{fig:n8decomp}
\end{minipage}
\end{figure}

Again we consider Dicke Model superradiance per \eq{eq:superradSOM}. For $N=8$ the superradiant populations are given by
\begin{align}\begin{split}\label{eq:8qubitsuperrad}
\chi _{0,8} &= e^{-8 \tau }\\
\chi _{1,7} &= \frac{4}{3} e^{-14 \tau } \left(e^{6 \tau }-1\right)\\
\chi _{2,6} &= \frac{1}{15} e^{-18 \tau } \left(-70 e^{4 \tau }+28 e^{10 \tau }+42\right)\\
\chi _{3,5} &= \frac{14}{5} e^{-20 \tau } \left(9 e^{2 \tau }-5 e^{6 \tau }+e^{12 \tau }-5\right)\\
\chi _{4,4} &= \frac{14}{3} e^{-20 \tau } \left(-60 \tau +54 e^{2 \tau }-10 e^{6 \tau }+e^{12 \tau }-45\right)\\
\chi _{5,3} &= \frac{28}{3} e^{-20 \tau } \left(75 (4 \tau +5)-25 e^{6 \tau }+e^{12 \tau }+27 e^{2 \tau } (20 \tau -13)\right)\\
\chi _{6,2} &= 28 e^{-20 \tau } \left(-50 e^{6 \tau } (3 \tau -2)+e^{12 \tau }-162 e^{2 \tau } (5 \tau -2)-25 (12 \tau +17)\right)\\
\chi _{7,1} &= \frac{196}{5} e^{-20 \tau } \left(125 e^{6 \tau } (2 \tau -1)+125 (2 \tau +3)+e^{12 \tau } (10 \tau -7)+81 e^{2 \tau } (10 \tau -3)\right)\\
\chi _{8,0} &= -800 e^{-14 \tau } (7 \tau -3)-196 e^{-20 \tau } (20 \tau +31)+\frac{49}{5} e^{-8 \tau } (23-40 \tau )+\frac{1568}{5} e^{-18 \tau } (11-45 \tau )+1
\end{split}\end{align}
which are plotted in \fig{fig:n8superad}.
Recall again that the decomposition parameters are solved from \eq{eq.qubitdecompSOM}. Enumerated explicitly for $N=8$ the decomposition equations are
\begin{align}\begin{split}
\chi _{0,8} &= x_1 \left(1-y_1\right){}^8+x_2 \left(1-y_2\right){}^8+x_3 \left(1-y_3\right){}^8+x_4 \left(1-y_4\right){}^8+x_5\\
\frac{\chi _{1,7}}{8} &= x_1 y_1 \left(1-y_1\right){}^7+x_2 \left(1-y_2\right){}^7 y_2+x_3 \left(1-y_3\right){}^7 y_3+x_4 \left(1-y_4\right){}^7 y_4\\
\frac{\chi _{2,6}}{28} &= x_1 y_1^2 \left(1-y_1\right){}^6+x_2 \left(1-y_2\right){}^6 y_2^2+x_3 \left(1-y_3\right){}^6 y_3^2+x_4 \left(1-y_4\right){}^6 y_4^2\\
\frac{\chi _{3,5}}{56} &= x_1 y_1^3 \left(1-y_1\right){}^5+x_2 \left(1-y_2\right){}^5 y_2^3+x_3 \left(1-y_3\right){}^5 y_3^3+x_4 \left(1-y_4\right){}^5 y_4^3\\
\frac{\chi _{4,4}}{70} &= x_1 \left(1-y_1\right){}^4 y_1^4+x_2 \left(1-y_2\right){}^4 y_2^4+x_3 \left(1-y_3\right){}^4 y_3^4+x_4 \left(1-y_4\right){}^4 y_4^4\\
\frac{\chi _{5,3}}{56} &= x_1 \left(1-y_1\right){}^3 y_1^5+x_2 \left(1-y_2\right){}^3 y_2^5+x_3 \left(1-y_3\right){}^3 y_3^5+x_4 \left(1-y_4\right){}^3 y_4^5\\
\frac{\chi _{6,2}}{28} &= x_1 \left(1-y_1\right){}^2 y_1^6+x_2 \left(1-y_2\right){}^2 y_2^6+x_3 \left(1-y_3\right){}^2 y_3^6+x_4 \left(1-y_4\right){}^2 y_4^6\\
\frac{\chi _{7,1}}{8} &= x_1 \left(1-y_1\right) y_1^7+x_2 \left(1-y_2\right) y_2^7+x_3 \left(1-y_3\right) y_3^7+x_4 \left(1-y_4\right) y_4^7\\
\chi _{8,0} &= x_1 y_1^8+x_2 y_2^8+x_3 y_3^8+x_4 y_4^8
\end{split}\end{align}
which we do not attempt to give an an analytic solution to. We stress that the system of equations defined by \eq{eq.qubitdecompSOM} is trivially enumerated for arbitrary $N$. Furthermore, most any program can solve the system of equations for numeric values of $\bchi$.

Since the system of equations is readily solvable  numerically, just as with $N=4$ we take the populations as governed by superradiance, now per \eqs{eq:8qubitsuperrad}, and for $N=8$ we restrict our consideration to numerical values of $\tau$. This restriction is entirely irrelevant, however, as our end-goal is to plot the decomposition parameters as (numeric) functions $\tau$. Doing so, we obtain \fig{fig:n8decomp} where again it is plainly evident that the extrama of $\brackets{\vec{x}(\tau),\vec{y}(\tau)}$ lie between zero and one. This visually certifies the perpetual separability of the system, and hence the inability of pure Dicke Model superradiance to generate entanglement. 

\FloatBarrier
\noindent\makebox[\linewidth]{\rule{\textwidth}{6pt}} 
\section{Complete Derivation of the SDS Form}
In the main text it is claimed that the definitions of the SDS form given in \eq{eq.qubitSDSrawdef} and \eq{eq.qubitSDS} are equivalent, meaning that
\begin{align}
	\int\limits_{0}^{2\pi}{ {\left(2 \pi\right)}^{-1}\sum\limits_{j=1}^{j_{\text{max}}}{x_j {{\left({\rho^1}\left[y_j,\phi\right]\right)}^{\otimes N}}\mbox{d}\phi}}
	= N!\sum\limits_{\bn}^{}{\sum\limits_{j=1}^{j_{\text{max}}}{{\frac{x_j {y_j}^{n_0}{(1-y_j)}^{n_1}}{n_0!n_1!}}\ket{D_\bn}\bra{D_\bn}}}
\end{align}
which we formally prove below.

\begin{enumerate}[leftmargin=0.5cm,itemindent=0cm,labelwidth=\itemindent,labelsep=0.5cm,align=parleft]
    \item  As in the main text preceding \eq{eq.qubitSDSrawdef} we take a completely generic normalized single-qubit pure state $\ket{\psi}\equiv \sqrt{y} \ket{0}+\sqrt{1-y} e^{\ic\phi}\ket{1}$ and use it to form a pure single-qubit product state, $\rho^1\left[y,\phi \right]\equiv \ket{\psi}\bra{\psi}$. Explicit expansion tells us that 
\begin{align}\begin{split}\label{eq:appendyform}
\rho^1\left[y,\phi \right]&=y \dyad{0}{0}+(1-y)\dyad{1}{1}+\sqrt{y(1-y)}(e^{-\ic\phi}\dyad{0}{1}+e^{\ic\phi}\dyad{1}{0})
\end{split}\end{align}
	\item Next take the tensor product of the single qubit product state with itself $N$ times, $\rho^N\left[y,\phi \right]\equiv\rho^1\left[y,\phi \right]^{\otimes N}$.  Raising a sum to a power $N$ results in a sum of products. Here the exponents $\gamma_{00},\gamma_{10},\gamma_{01},\gamma_{11}$ appearing in the products below are to be understood as ranging over nonnegative integers $\bm{\gamma}\in \mathbb{Z}^+$ in such a manner that the sum of the exponents total $N$, $\gamma_{00}+\gamma_{10}+\gamma_{01}+\gamma_{11}=N$.
\begin{align}\begin{split}
    \rho^N\left[y,\phi \right]\equiv&\rho^1\left[y,\phi \right]^{\otimes N}
    \\=&\sum_{\lbrace{\text{all }\gamma \rbrace}}^N{
    y^{(\gamma_{00}+\gamma_{01}/2)}(1-y)^{(\gamma_{11}+\gamma_{10}/2)}e^{\ic\phi(\gamma_{10}-\gamma_{01})}\sum\limits_{\begin{subarray}{l} \text{operator permutations} \\
\brackets{\dyad{0}{0},\dyad{0}{1},\dyad{1}{0},\dyad{1}{1}} \end{subarray}}{
    \rho\begin{bmatrix}\gamma_{00} & \gamma_{01} \\ \gamma_{10} & \gamma_{11} \end{bmatrix}}}
\end{split}\end{align}
where we have introduced a convenient generalization of computational basis states for product states,
\begin{align}
\rho\begin{bmatrix}\gamma_{00} & \gamma_{01} \\ \gamma_{10} & \gamma_{11} \end{bmatrix}\equiv \pdyad{0}{0}^{\otimes \gamma_{00}}\pdyad{1}{1}^{\otimes \gamma_{11}}\pdyad{0}{1}^{\otimes \gamma_{01}}\pdyad{1}{0}^{\otimes \gamma_{10}}\,.
\end{align}
Note that the sum over operator permutations is intentionally {\em not} normalized as each permutation of each has equal weight in the expansion of $\rho^1\left[y,\phi \right]^{\otimes N}$. 
	\item  The next step is to mix uniformly over all $\phi$, namely $\rho^N\left[y\right]\equiv \left(2\pi\right)^{-1}\int_{0}^{2\pi}{\rho^N\left[y,\phi\right]\;\mbox{d}\phi}$ . The trick in this step is that
\begin{align}
\int\limits_{0}^{2\pi} e^{\ic \phi\left(\gamma_{10}-\gamma_{01}\right)} \D \phi = \begin{cases} 0 & \gamma_{10} \neq \gamma_{01} \\ 1 & \gamma_{10} = \gamma_{01}\end{cases}
\end{align}
which allows us to perform a change-of-variable such that $\gamma_{10}= \gamma_{01}\rightarrow \kappa,\;\; \gamma_{00} \rightarrow n_0-\kappa,\;\; \gamma_{11} \rightarrow n_1-\kappa$, yielding simply
\begin{align}
\rho^N\left[y\right]&= \sum\limits_{\bn}^{}{\sum\limits_{\kappa}^{}{{y}^{n_0}{(1-y)}^{n_1}\sum\limits_{\begin{subarray}{l} \text{operator permutations} \\
\brackets{\dyad{0}{0},\dyad{0}{1},\dyad{1}{0},\dyad{1}{1}} \end{subarray}}{
    \rho\begin{bmatrix}\left(n_0 - \kappa\right) & \kappa \\ \kappa, & \left(n_1 -\kappa\right) \end{bmatrix}}}}      
\end{align}
where instead of summing over the four $\gamma$'s we are summing over $n_0,n_1$, and $\kappa$. In these variables new the condition $\gamma_{00}+\gamma_{10}+\gamma_{01}+\gamma_{11}=N$ is automatically satisfied, but to preserve the positivity of both $\gamma_{00}$ and $\gamma_{11}$ we must be careful to upper bound $\kappa \leq \operatorname{min}[n_0,n_1]$.
\item To proceed we must notice that 
\begin{align}\label{eq:counting}
\sum\limits_{\begin{subarray}{l} \text{operator permutations} \\
\brackets{\dyad{0}{0},\dyad{0}{1},\dyad{1}{0},\dyad{1}{1}} \end{subarray}} {\hspace{-1em}\sum\limits_{\kappa}^{\operatorname{min}[n_0,n_1]}{
    \rho\begin{bmatrix}\left(n_0 - \kappa\right) & \kappa \\ \kappa, & \left(n_1 -\kappa\right) \end{bmatrix}}} = \begin{pmatrix}\sum\limits_{\begin{array}{c}\scriptstyle{\text{perms.}}\\ \scriptstyle{\brackets{\ket{0},\ket{1}}}\end{array}}{\ket{\underbrace{0...0}_{n_0}\underbrace{1...1}_{n_1}}}\end{pmatrix}\begin{pmatrix}\sum\limits_{\begin{array}{c}\scriptstyle{\text{perms.}}\\ \scriptstyle{\brackets{\bra{0},\bra{1}}}\end{array}}{\bra{\underbrace{0...0}_{n_0}\underbrace{1...1}_{n_1}}}\end{pmatrix}
\end{align}
which makes use of a binomial theorem argument. The left hand side of \eq{eq:counting} is a double sum, over permutations of the four operators as well as over all possible partition schemes indexed by $k$. This is equivalent to the right hand side of \eq{eq:counting}, namely taking the product of unpaired permutation summations. This counting scheme follows from $\sum_{\kappa}{\frac{N!}{\kappa!(n_0-\kappa)!(n_1-\kappa)!\kappa!}}={\left(\frac{N!}{n_0!n_1!}\right)}^2\,$. As an explicit example consider the sixteen terms of
\begin{align}\begin{split}\label{eq:specialcaseSOM}
&\sum\limits_{\begin{subarray}{l} \text{operator permutations} \\\brackets{\dyad{0}{0},\dyad{0}{1},\dyad{1}{0},\dyad{1}{1}} \end{subarray}} {\hspace{-1em}\sum\limits_{\kappa}^{\operatorname{min}[n_0,n_1]}{\rho\begin{bmatrix}\left(3 - \kappa\right) & \kappa \\ \kappa, & \left(1 -\kappa\right) \end{bmatrix}}} 
=\sum\limits_{\begin{subarray}{l} \text{operator permutations} \\\brackets{\dyad{0}{0},\dyad{0}{1},\dyad{1}{0},\dyad{1}{1}} \end{subarray}} {\left(\rho\begin{bmatrix} 3 & 0 \\ 0, & 1 \end{bmatrix}+\rho\begin{bmatrix} 2 & 1 \\ 1, & 0 \end{bmatrix}\right)} 
\\ &\,=\hphantom{}\Big(\op{\da}{\da}{\da}{\dd}+\op{\da}{\da}{\dd}{\da}+\op{\da}{\dd}{\da}{\da}+\op{\dd}{\da}{\da}{\da}\Big)
\\&\,\hphantom{=}+\Big(\op{\da}{\da}{\db}{\dc}+\op{\da}{\da}{\dc}{\db}+\op{\da}{\db}{\da}{\dc}+\op{\da}{\dc}{\da}{\db}
\\&\,\hphantom{=\Big(\Big(}+\op{\da}{\db}{\dc}{\da}+\op{\da}{\dc}{\db}{\da}+\op{\db}{\da}{\da}{\dc}+{\dc}{\da}{\da}{\db}
\\&\,\hphantom{=\Big(\Big(}+\op{\db}{\da}{\dc}{\da}+\op{\dc}{\da}{\db}{\da}+\op{\db}{\dc}{\da}{\da}+\op{\dc}{\db}{\da}{\da}\Big)
\\ &\,= \hphantom{}\Big(\ket{0001}+\ket{0010}+\ket{0100}+\ket{1000}\Big) \Big(\bra{0001}+\bra{0010}+\bra{0100}+\bra{1000}\Big)\,.
\end{split}\end{align}

At this point it is constructive to review \eq{eq:firstdef} in the main text, which defined the Dicke states as ${\ket{D_\bn} =\nobreak w_\bn \sum_{\begin{array}{c}\scriptstyle{\text{perms.}}\\ \scriptstyle{\brackets{\ket{0},\ket{1}}}\end{array}}{\ket{\underbrace{0...0}_{n_0},\underbrace{1...1}_{n_1}}}}$. Note that the special case of $\bn=\brackets{3,1}$ considered in \eq{eq:specialcaseSOM} is the identically the example of \eq{eq:exampleket} in the main text. Making use of \eq{eq:counting} with \eq{eq:firstdef} allows for a direct substitution such that we have
\begin{align}
\rho^N\left[y\right]&= \sum\limits_{\bn}^{}{\frac{{y}^{n_0}{(1-y)}^{n_1}}{{w_\bn}^2}\ket{D_\bn}\bra{D_\bn}}\,.
\end{align}
	\item The last step in our construction is to take an arbitrary finite convex mixture over multiple possible $y_j$ so that each $\rho^N\left[y_j\right]$ gets weighted by some parameter $x_j$, 
    $\rho_{\text{\tiny SDS}}\equiv\sum_{j=1}^{j_{\text{max}}}{x_j \rho^N\left[y_j\right]}$. We substitute in for the definition of $w_\bn=\sqrt{n_0!n_1!/N!}$ to finally match up with the quoted form of \eq{eq.qubitSDS} in the main text,
	\begin{align}\label{eq.qubitSDS.SOM}
\rho_{\text{\tiny SDS}}&=N!\sum\limits_{\bn}^{}{\sum\limits_{j=1}^{j_{\text{max}}}{{\frac{x_j {y_j}^{n_0}{(1-y_j)}^{n_1}}{n_0!n_1!}}\ket{D_\bn}\bra{D_\bn}}}
\end{align}
thereby proving that
\begin{align}
\int\limits_{0}^{2\pi}{{\left(2\pi\right)}^{-1}\sum\limits_{j=1}^{j_{\text{max}}}{x_j {\rho\left[y_j,\phi\right]^{\otimes N}}\mbox{d}\phi}}=N!\sum\limits_{\bn}^{}{\sum\limits_{j=1}^{j_{\text{max}}}{{\frac{x_j {y_j}^{n_0}{(1-y_j)}^{n_1}}{n_0!n_1!}}\ket{D_\bn}\bra{D_\bn}}}
\end{align}
as claimed.
\end{enumerate}
For completeness, recall that we define $j_{\text{max}}=\lceil\left(N+1\right)/2\rceil$ with the special restriction such that $y_{\left(N+2\right)/2}=0$ when $N$ is even, per the discussion subsequent to \eq{eq.qubitdecomp} in the main text.

\vspace{5ex}
\noindent\makebox[\linewidth]{\rule{\textwidth}{6pt}} 
\section{Volume of the Separable States for arbitrary N}
The volume calculations to determine $\operatorname{PPTGDSVol}$ and $\operatorname{SDSVol}$ can be done (easily and analytically) for the trivial cases of $N=2,3$  and indeed we find perfect {\em analytic} agreement between the $\operatorname{PPTGDSVol}_{N}$ and the $\operatorname{SDSVol}_{N}$ for those cases. Such small-$N$ considerations are useful only insofar as verifying the methodology, as the PPT criterion is known to be necessary and sufficient for separability in those regimes \cite{ppt4qubit,Lowenstein2012PPT}.

It is interesting to consider larger $N$ however, for which $\operatorname{PPTGDSVol}_{N}$, the generalization of \eq{GDSVolIntegral} from the main text, becomes computationally intractable. On the other hand $\operatorname{SDSVol}_{N}$ can be readily calculated analytically up through $N\! \sim \Or{(10)}$, as its discontinuous indicator function appearing in the integrand is much simpler. For $\operatorname{SDSVol}_{N}$ the purpose of the indicator function is merely to ensure a one-to-one mapping between $\bchi$ and $\bx,\by$, and it can be substituted for nothing more than division by the multiplicity of solutions to the polynomial system of equations produced by \eq{eq.qubitdecomp} of the main text, leaving the integrand as just a lone volume element. We tabulated $\operatorname{SDSVol}_{N}$ for many $N$ and found that it fits the formula
\begin{align}\label{NSDSVol}
\operatorname{SDSVol}_{N} = \prod_{z=1}^{N}{z^{(z-1)}\frac{(n-1)!}{(2 n-1)!}}
\end{align}
although we have not yet been able to derive this from first principles. \eq{NSDSVol} provably yields the volume of the separable GDS states for $N\leq 4$, as for those cases we thoroughly demonstrated that $\brho_{\mbox{\tiny SDS}} = \brho_{\mbox{\tiny SEP} \cap \mbox{\tiny GDS}}$. If one also accepts that Lemma (\ref{PPTSEP}) of the main text holds true for all $N$ then one has that \eq{NSDSVol} yields the volume of the separable GDS states for all $N$.
\end{document}